\documentclass[twocolumn,preprintnumbers,amsmath,amssymb,superscriptaddress]{revtex4}
\usepackage{graphicx}
\usepackage{dcolumn}
\usepackage{bm}
\usepackage{soul}
\usepackage{color}
\usepackage{epstopdf}
\usepackage[version=3]{mhchem}
\usepackage{lipsum}
\usepackage[outercaption]{sidecap}
\usepackage{floatrow}
\usepackage{hyperref}
\usepackage{appendix}

\begin{document}

\title{Theoretical Model for the High-Pressure Melting Process\\ of MgO with the B1 Structure}
\author{Tran Dinh Cuong}
\email{cuong.trandinh@phenikaa-uni.edu.vn}
\affiliation{Faculty of Materials Science and Engineering, Phenikaa University, Hanoi 12116, Vietnam}
\author{Anh D. Phan}
\email{anh.phanduc@phenikaa-uni.edu.vn}
\affiliation{Faculty of Materials Science and Engineering, Phenikaa University, Hanoi 12116, Vietnam}
\affiliation{Phenikaa Institute for Advanced Study (PIAS), Phenikaa University, Hanoi 12116, Vietnam}
\date{\today}

\date{\today}

\begin{abstract}
MgO is an abundant mineral in the rocky mantle of terrestrial planets, but its melting behaviors remain enigmatic. Here we introduce a simple theoretical model to investigate the B1-liquid transition of MgO up to 370 GPa. Vibrational free energies of B1-MgO are fully computed by the moment recurrence technique in quantum statistical physics. On that basis, we associate the melting temperature with the isothermal bulk modulus via the work-heat equivalence principle. This strategy allows us to quantitatively explain recent experimental data. Our numerical analyses would yield insights into planetary dynamics and evolution.
\end{abstract}

\maketitle
\section{Introduction}
The melting transition of MgO has garnered tremendous attention for decades \cite{1}. This oxide is widely used in steel metallurgy \cite{2}, chemical engineering \cite{3}, nuclear technology \cite{4}, solar energy applications \cite{5}, etc., due to its superior refractoriness. For Earth sciences, the melting properties of MgO play a crucial role in explaining the rheology \cite{6}, heterogeneity \cite{7}, and seismic-velocity anomaly \cite{8} of the lower mantle. Furthermore, recent first-principles studies \cite{9} have revealed the abundance of MgO in solar giants and terrestrial exoplanets. Consequently, their basic thermochemical state can be determined via the melting diagram of MgO at ultra-high pressure \cite{10}. 

Conventionally, one can investigate the melting phenomenon by diamond anvil cell (DAC) techniques \cite{11}, shock-wave (SW) methods \cite{12}, or \textit{ab initio} molecular dynamics (AIMD) simulations \cite{13}. However, an alarming discrepancy among these approaches has arisen in a variety of refractory materials \cite{14,15,16,17,18}. At 135 GPa, the melting temperature of MgO spans from 5000 to 9000 K \cite{19}. Hence, phase relations at the Earth's core-mantle boundary are not well constrained \cite{20}. In addition, running AIMD codes requires enormous computational resources \cite{21}. For strongly ionic systems, the simulation complexity rises dramatically because of the high plane-wave cutoff energy \cite{22}. Despite numerous attempts to resolve these problems, a unified picture of molten MgO has not been reached.  

Lately, the statistical moment method (SMM) \cite{23,24,25,26,27,28} has been developed to gain insights into melting mechanisms under extreme conditions. The SMM directly links the average atomic displacement and its high-order counterparts to estimate anharmonic free energies \cite{23,24,25}. On that basis, one can deduce the melting boundary from macroscopic information of solid phases (e.g., equation of state and elastic moduli) \cite{26,27,28}. This scheme helps reproduce experimental data with minimal computational costs \cite{26,27,28}. Notwithstanding, most SMM analyses are restricted to pure metals \cite{26,27} and solid solutions \cite{28}.  

Here we extend the SMM model \cite{23,24,25,26,27,28} to capture the melting behaviors of MgO in planetary interiors. Theoretical calculations are performed for the B1 structure in a range of 0 to 370 GPa \cite{29}. Our numerical results are systematically compared with prior works. 

\section{Theoretical background}
In the SMM \cite{23,24,25,26,27,28}, an arbitrary $X$ ion ($X=$ Mg or O) is described by three quantum oscillators having the Hooke constant $k_X$ and nonlinear parameters $\gamma_{1X}$, $\gamma_{2X}$, $\gamma_X$. Applying the Leibfried-Ludwig lattice theory \cite{30} to the cohesive energy $u_{0X}$ gives
\begin{eqnarray}
k_X&=&\frac{1}{2}\sum_i\left(\frac{\partial^2u_{0X}}{\partial u_{i\alpha}^2}\right)_{eq},\;\gamma_{1X}=\frac{1}{48}\sum_i\left(\frac{\partial^4u_{0X}}{\partial u_{i\alpha}^4}\right)_{eq},\nonumber\\
\gamma_{2X}&=&\frac{1}{8}\sum_i\left(\frac{\partial^4u_{0X}}{\partial u_{i\alpha}^2\partial u_{i\beta}^2}\right)_{eq},\;\gamma_X=4\left(\gamma_{1X}+\gamma_{2X}\right),\nonumber\\
\alpha&\ne&\beta=x,y,z,
\label{eq:1}
\end{eqnarray}
where $u_{i\alpha}$ and $u_{i\beta}$ denote the displacement of other ions along Cartesian axes. From these structural properties, the vibrational free energy $F_X$ is determined by \cite{23,24,25}
\begin{eqnarray}
&F_X&(a_X,T)=\frac{1}{2}u_{0X}+3\theta\left[\zeta_X+\ln\left(1-e^{-2\zeta_X}\right)\right]\nonumber\\
&+&\frac{3\theta^2}{k_X^2}\left[\gamma_{2X}\eta_X^2-\frac{2}{3}\gamma_{1X}\left(1+\frac{1}{2}\eta_X\right)\right]\nonumber\\
&+&\frac{8\theta^3}{k_X^4}\gamma_{2X}^2\eta_X\left(1+\frac{1}{2}\eta_X\right)\nonumber\\
&-&\frac{12\theta^3}{k_X^4}\gamma_{1X}\left(\gamma_{1X}+2\gamma_{2X}\right)\left(1+\eta_X\right)\left(1+\frac{1}{2}\eta_X\right),\nonumber\\
\theta&=&k_BT,\quad\zeta_X=\frac{\hbar\omega_X}{2\theta},\quad\eta_X=\zeta_X\coth\zeta_X,
\label{eq:2}
\end{eqnarray}
where $a_X$ is the nearest neighbor distance, $T$ is the absolute temperature, $k_B$ is the Boltzmann constant, $\hbar$ is the reduced Planck constant, and $\omega_X$ is close to the Einstein frequency.

The thermodynamic definition of pressure is
\begin{eqnarray}
P=-\frac{1}{3a_X^3}\left(\frac{\partial F_X}{\partial a_X}\right)_T.
\label{eq:3}
\end{eqnarray}
At $T=0$ K, since anharmonic contributions are negligible, one can simplify Eq.(\ref{eq:3}) by \cite{26}
\begin{eqnarray}
P=-\frac{1}{2a_X^2}\left(\frac{1}{3}\frac{\partial u_{0X}}{\partial a_X}+\frac{\hbar\omega_X}{2k_X}\frac{\partial k_X}{\partial a_X}\right).
\label{eq:4}
\end{eqnarray}
Solving the cold equation of state provides specific values of $a_X(P,0)$. When $T>0$ K, the ionic arrangement is characterized by \cite{26,27}
\begin{eqnarray}
a_X(P,T)=a_X(P,0)+y_X(P,T),
\label{eq:5}
\end{eqnarray}
\begin{eqnarray}
y_X(P,T)=\sqrt{\frac{2\gamma_X\theta^2}{3k_X^3}A_X},
\label{eq:6}
\end{eqnarray}
where $y_X(P,T)$ indicates the lattice dilation upon heating. The analytical expression for $A_X$ was established in Ref.\cite{23} via force-balance conditions. Based on Eq.(\ref{eq:5}), the bulk modulus $K_T$ of MgO is approximated by \cite{28}
\begin{eqnarray}
K_T=-\sum_{X}\frac{c_X}{3}\frac{a_X^3(P,0)}{a_X^2(P,T)}\left(\frac{\partial P}{\partial a_X}\right)_T,
\label{eq:7}
\end{eqnarray}
where $c_X=0.5$ is the ionic concentration.

It is worth noting that elastic responses have an intimate correlation with the solid-liquid transition \cite{31,32,33}. According to Ma \textit{et al.} \cite{31}, a liquid sample can be solidified due to exothermal effects or compression. Thus, heat energies upon isobaric cooling are equivalent to mechanical works upon isothermal squeezing \cite{31}. This physical picture gives us the explicit form of the melting temperature as \cite{31}
\begin{widetext}
\begin{eqnarray}
T_m(P)&=&T_m(0)+\left(T_m(P^*)-T_m(0)\right)\sqrt{\cfrac{\left(1+\cfrac{P}{K_T(0)}\right)\left(\cfrac{K_T(P)}{K_T(0)}\right)^{-1/K_T^{'}(0)}-1}{\left(1+\cfrac{P^*}{K_T(0)}\right)\left(\cfrac{K_T(P^*)}{K_T(0)}\right)^{-1/K_T^{'}(0)}-1}},
\label{eq:8}
\end{eqnarray}
\end{widetext}
where $P^*>0$ GPa is an arbitrarily fixed pressure and $K_T^{'}=\left(\partial K_T/\partial P\right)_T$.

Equation (\ref{eq:8}) can accurately regenerate experimental data for various substances via two reference melting points $T_m(0)$ and $T_m(P^*)$ \cite{31}. Unfortunately, in the case of MgO \cite{1}, melting behaviors at $P^*$ are very elusive. Even if $P^*=3$ GPa, the difference among static measurements can be up to 20 \% \cite{19}. To handle this issue, we reconsider Eq.(\ref{eq:8}) in the limit $P^*\to0$ GPa, which is
\begin{widetext}
\begin{eqnarray}
T_m(P)=T_m(0)+\left(\frac{dT_m}{dP}\right)_{P=0}\sqrt{\cfrac{2K_T^2(0)\left[\left(1+\cfrac{P}{K_T(0)}\right)\left(\cfrac{K_T(P)}{K_T(0)}\right)^{-1/K_T^{'}(0)}-1\right]}{K_T^{'}(0)-1}}.
\label{eq:9}
\end{eqnarray}
\end{widetext}
Remarkably, Ma \textit{et al.} have revealed a good agreement between the work-heat equivalence principle (WHEP) \cite{31} and the dislocation-mediated melting theory \cite{32} at low pressure. Therefore, we can connect the initial melting gradient with elastic properties by \cite{32} 
\begin{eqnarray}
\left(\frac{dT_m}{dP}\right)_{P=0}=T_m(0)\left(\frac{G^{'}(0)}{G(0)}-\frac{1}{K_T(0)}\right),
\label{eq:10}
\end{eqnarray}
where $G$ and $G'=\left(\partial G/\partial P\right)_T$ are the shear modulus and its pressure derivative, respectively. If the Poisson ratio is insensitive to compression, Eq.(\ref{eq:10}) is rewritten by \cite{32}
\begin{eqnarray}
\left(\frac{dT_m}{dP}\right)_{P=0}=T_m(0)\frac{K_T^{'}(0)-1}{K_T(0)}.
\label{eq:11}
\end{eqnarray}
The obtained result is quantitatively consistent with the Lindemann criterion for vibrational instability \cite{33}. Inserting Eq.(\ref{eq:11}) into Eq.(\ref{eq:9}), we have
\begin{widetext}
\begin{eqnarray}
T_m(P)=T_m(0)\left\{1+\sqrt{2\left(K_T^{'}(0)-1\right)\left[\left(1+\frac{P}{K_T(0)}\right)\left(\frac{K_T(P)}{K_T(0)}\right)^{-1/K_T^{'}(0)}-1\right]}\right\}. 
\label{eq:12}
\end{eqnarray}
\end{widetext}
It is clear to see that Eq.(\ref{eq:12}) only requires an experimental value of $T_m(0)$. This modification makes the WHEP model \cite{31} more predictive. The effectiveness of Eq.(\ref{eq:12}) is numerically demonstrated in the Supplementary Material. In the next section, we combine Eq.(\ref{eq:7}) with Eq.(\ref{eq:12}) to access the melting boundary of MgO.

\section{RESULTS AND DISCUSSION}

To describe the interionic interaction in oxide compounds, one often adopts the Born-Mayer-Buckingham pair potential as \cite{34,35}
\begin{eqnarray}
\varphi_{ij}\left(r_{ij}\right)=B_{ij}\exp\left(-\frac{r_{ij}}{\rho_{ij}}\right)-\frac{C_{ij}}{r_{ij}^6}+\frac{q_iq_j}{4\pi\epsilon_0r_{ij}}f\left(r_{ij}\right),
\label{eq:13}
\end{eqnarray}
where $r_{ij}$ is the separation between ions $i$ and $j$, $B_{ij}$ and $C_{ij}$ are repulsive and attractive constants, $\rho_{ij}$ is the decay length, $q_i$ and $q_j$ are effective charges, $\epsilon_0$ is the vacuum permittivity, and $f\left(r_{ij}\right)$ is a short-range function. Utilizing the damped shifted force method yields \cite{36}
\begin{widetext}
\begin{eqnarray}
f\left(r_{ij}\right)=\begin{cases}\textrm{erfc}\left(\alpha^{*} r_{ij}\right)-\cfrac{\textrm{erfc}\left(\alpha^{*} R_c\right)}{R_c}r_{ij}+\left(\cfrac{\textrm{erfc}\left(\alpha^{*} R_c\right)}{R_c^2}+\cfrac{2\alpha^{*}}{\sqrt{\pi}}\cfrac{\exp\left(-\alpha^{*2}R_c^2\right)}{R_c}\right)r_{ij}\left(r_{ij}-R_c\right), &\mbox{$r_{ij}\leq R_c$}\\\qquad\qquad\qquad\qquad\qquad\qquad\qquad\qquad\qquad\qquad\qquad\qquad\qquad\qquad\qquad\qquad\qquad\;\,\,0,&\mbox{$r_{ij}>R_c$}\end{cases}
\label{eq:14}
\end{eqnarray}
\end{widetext}
where $\alpha^{*}$ is the damping factor \cite{37} and $R_c$ is the cutoff radius \cite{38}. The Born-Mayer-Buckingham parameters for MgO are presented in Table \ref{tab:table1} by applying the Pareto optimization to machine learning \cite{34}. 

\begin{table}[htp]
\begin{ruledtabular}
\begin{tabular}{cccc}
\textrm{Bond}&
\textrm{$B_{ij}$ (eV)}&
\textrm{$C_{ij}$ (eV.\AA$^6$)}&
\textrm{$\rho_{ij}$ (\AA)}\\
\colrule
Mg$^{+1.7}$ - Mg$^{+1.7}$&0 &0 &0.5
\\
O$^{-1.7}$ - O$^{-1.7}$&23232.83& 57.45&0.22658
\\
Mg$^{+1.7}$ - O$^{-1.7}$&1248.66 & 0 &0.29113
\end{tabular}
\caption{\label{tab:table1} The Born-Mayer-Buckingham parameters for MgO taken from machine learning algorithms \cite{34}. Here we employ $\alpha^{*}=0.2$ \AA$^{-1}$ \cite{37} and $R_c=10.525$ \AA\ \cite{38} to treat long-range electrostatic interactions.}
\end{ruledtabular}
\end{table}

\begin{figure}[htp]
\includegraphics[width=9 cm]{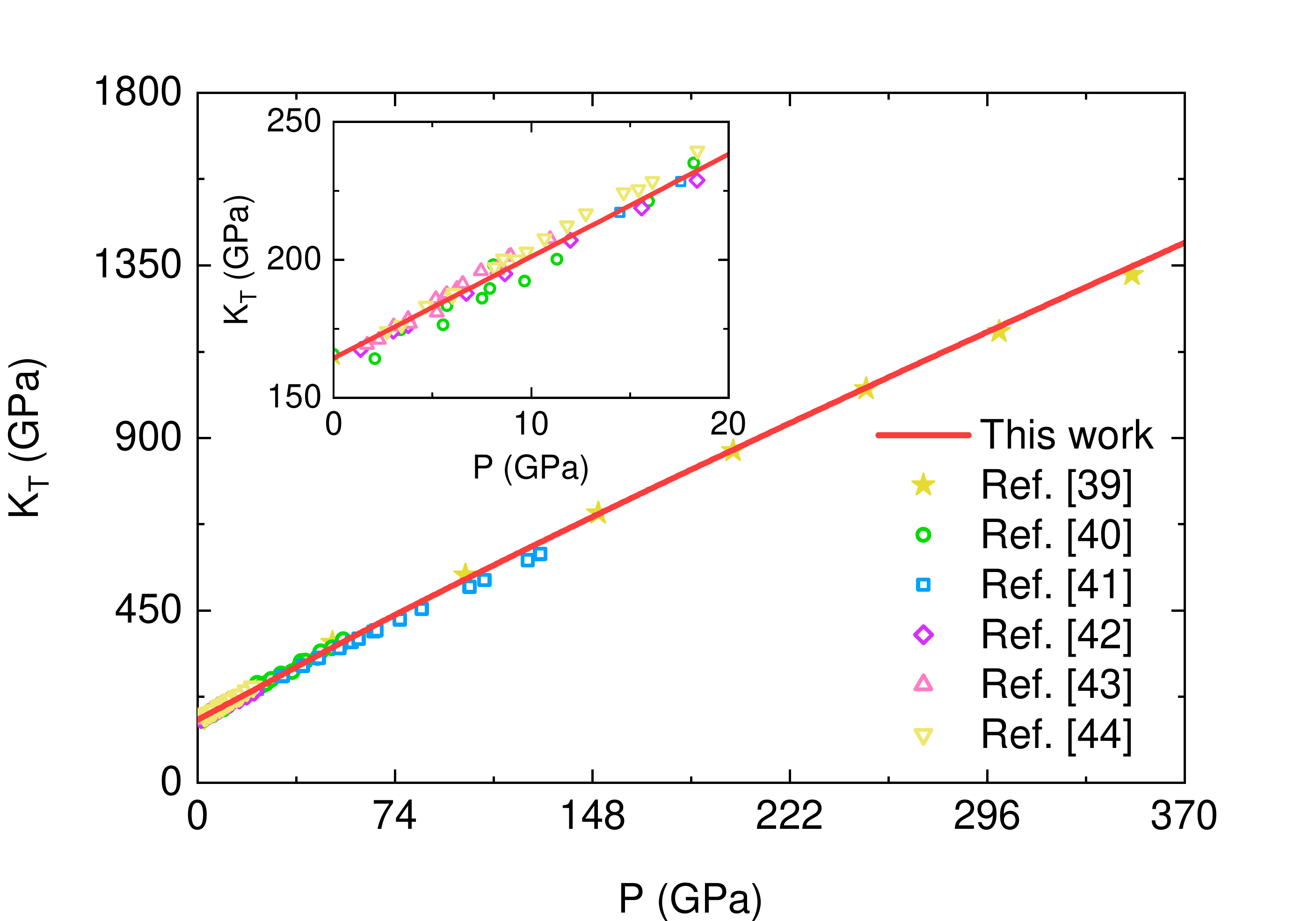}
\caption{\label{fig:1}(Color online) Correlations between pressure and the bulk modulus of MgO derived from Eq.(\ref{eq:7}), first-principles calculations \cite{39}, and experiments \cite{40,41,42,43,44}.}
\end{figure}

Figure \ref{fig:1} shows how the bulk modulus of MgO depends on pressure at room temperature. One can realize that $K_T$ increases almost linearly upon isothermal squeezing. Under ambient conditions, its growth rate is about $K_T^{'}=4.30$. SMM analyses only deviate from first-principles calculations \cite{39}, Brillouin scattering experiments \cite{40,41,42}, and ultrasonic measurements \cite{43,44} by a few percent. This excellent accordance affirms our theoretical approach and the chosen interionic potential. In practice, precise knowledge of $K_T$ and $K_T^{'}$ is useful for modeling planetary interiors \cite{45}, designing advanced materials \cite{46}, and developing high-pressure metrologies \cite{47}.

Figure 2 shows compression effects on the melting temperature of MgO. It is conspicuous that available results are in stark contrast with each other. Pioneering DAC works \cite{48} have reported an anomalously flat melting curve for MgO. At 0 GPa, the DAC melting slope is about 36 K.GPa$^{-1}$ \cite{48}, far below theoretical expectations \cite{49} and multi-anvil extrapolations \cite{50}. According to Aguado and Madden, this strange tendency may arise from thermal stress \cite{51}, surface melting \cite{52}, or solid-solid transition \cite{53}. Notably, Kimura \textit{et al.} \cite{54} have pointed out the confusion between ``melting" and ``plastic deformation" in previous DAC measurements \cite{48}. Based on the micro-texture analysis of quenched samples, Kimura \textit{et al.} \cite{54} have obtained significant growth in the melting point of MgO. Their findings \cite{54} have been validated by recent DAC studies on (Mg,Fe)O solid solutions \cite{55,56}. Nevertheless, all of these static experiments \cite{54,55,56} are limited to 120 GPa.  

For higher-pressure regimes, dynamic SW techniques are required \cite{12}. Fat’yanov \textit{et al.} have carried out subtle gas-gun measurements on MgO preheated to 1850 K \cite{19}. Surprisingly, no melting signatures have been detected along the Hugoniot up to 248 GPa and 9100 K \cite{19}. This lower bound for $T_m$ has rigorously ruled out theoretical interpretations in Ref.\cite{53,57,58,59}. 

From a computational perspective, using AIMD \cite{13} is an efficient strategy to explore deep-planetary interiors. This powerful tool can predict the melting behaviors of MgO without adjustable parameters \cite{60,61,62,63,64,65,66}. Notwithstanding, along the melting boundary, phase relations of MgO remain controversial \cite{60,61,62,63,64,65,66}. Boates and Bonev have suggested that MgO converts from B1 into B2 structures at 364 GPa and 12000 K \cite{62}. Unfortunately, their transition temperature \cite{62} is overestimated due to the inaccurate treatment of liquid entropies \cite{66}. On the other hand, Root \textit{et al.} \cite{63} have reported that MgO only melts from the B1 phase when $0\leq P\leq265$ GPa. This conclusion \cite{63} is well supported by most AIMD simulations \cite{64,65,66}. However, the lack of anharmonicity can lead to the underestimation of transition pressure \cite{67}. Remarkably, in the latest work \cite{29}, Soubiran and Militzer have successfully reconciled the contradiction among AIMD melting profiles \cite{60,61,62,63,64,65,66}. Relying on thermodynamic integrations, Soubiran and Militzer \cite{29} have found the B1-B2-liquid triple point to lie at 370 GPa and 10000 K. Their results \cite{29} are compatible with TDEP calculations \cite{67} and ramp compression experiments \cite{68}.

\begin{figure}[htp]
\includegraphics[width=9 cm]{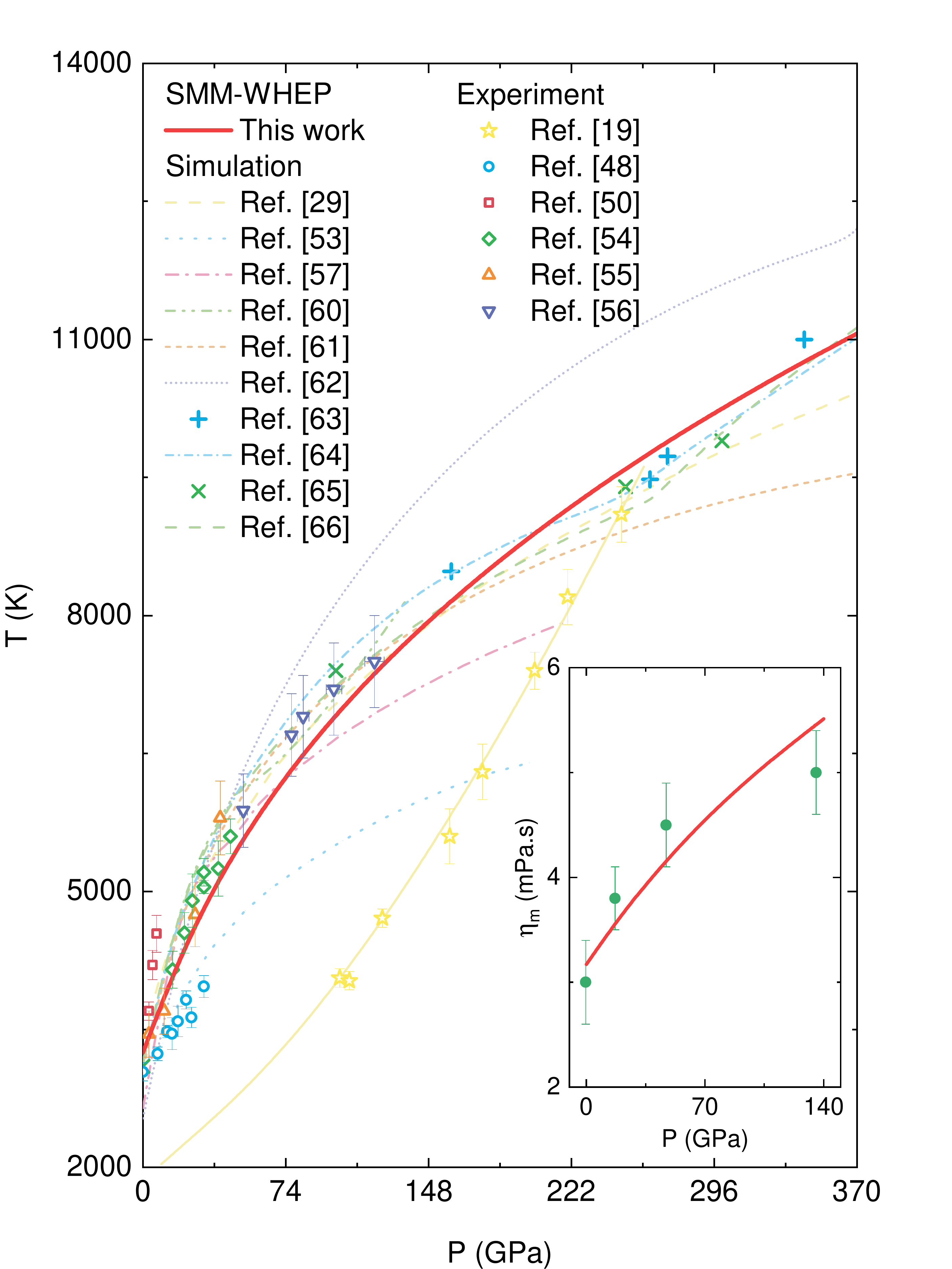}
\caption{\label{fig:2}(Color online) The high-pressure melting curve of MgO given by Eq.(\ref{eq:12}), experiments \cite{19,48,50,54,55,56}, and simulations \cite{29,53,57,60,61,62,63,64,65,66}. Note that the yellow solid line represents the Hugoniot in Ref.\cite{19}. Inset: The melt viscosity of MgO as a function of pressure inferred from Eq.(\ref{eq:16}) (red solid line) and AIMD computations \cite{60} (green filled point).}
\end{figure}

In general, we can quickly reconstruct high-quality datasets in Ref.\cite{29,54,55,56} via basic elastic information at room temperature. The discrepancy between the SMM-WHEP and other precise methods \cite{29,54,55,56} is less than 6 $\%$ in a range of 0 to 370 GPa. This value confirms the reliability of Eq.(\ref{eq:12}) for high-pressure melting processes. Besides, SMM-WHEP analyses satisfy tight SW criteria of Fat’yanov \textit{et al.} \cite{19}. Consequently, the real melting line of MgO may be very close to our predictions. 

To facilitate geodynamic models, we fit SMM-WHEP results by the Simon-Glatzel equation as \cite{69}
\begin{eqnarray}
T_m=T_{0}\left(\frac{P}{P_0}+1\right)^{1/c},
\label{eq:15}
\end{eqnarray}
where $T_0=3214$ K, $P_0=16.62$ GPa, and $c=2.54$. On that basis, the melt viscosity of MgO can be approximated by \cite{70}
\begin{eqnarray}
\eta_m(P)\propto\frac{\sum_Xc_X\omega_X\left(P,T_m\right)}{\sum_Xc_Xa_X\left(P,T_m\right)}.
\label{eq:16}
\end{eqnarray}
As shown in the inset, $\eta_m(P)$ increases gradually from 3.17 mPa.s at 0 GPa to 5.51 mPa.s at 140 GPa. Our numerical calculations agree quantitatively well with recent AIMD simulations \cite{60}. Geophysically, the low viscosity of molten MgO indicates turbulent convection and rapid cooling of early magma oceans on Earth \cite{71}. Thus, our new constraints on melting profiles would enhance understanding of the thermochemical evolution of terrestrial planets \cite{71}.

\section{CONCLUSION}
We have developed the SMM-WHEP model to fully describe the melting transition of B1-MgO up to 370 GPa. This approach allows us to directly determine high-pressure melting properties from equation-of-state parameters. Therefore, we can quantitatively interpret DAC, SW, and AIMD results without heavy computational workloads. Our findings would have profound geodynamic implications for the evolution of rocky planets. It is possible to extend our SMM-WHEP method to capture the melting behaviors of other minerals. 

\begin{acknowledgements}
T. D. Cuong is deeply grateful for the research assistantship at Phenikaa University. This research was funded by the Vietnam National Foundation for Science and Technology Development (NAFOSTED) under grant number 103.01-2019.318.
\end{acknowledgements}

\end{document}


\title{Supplementary Material for\\ Theoretical Model for the High-Pressure Melting Process \\of MgO with the B1 Structure}
\author{Tran Dinh Cuong}
\email{cuong.trandinh@phenikaa-uni.edu.vn}
\affiliation{Faculty of Materials Science and Engineering, Phenikaa University, Hanoi 12116, Vietnam}
\author{Anh D. Phan}
\email{anh.phanduc@phenikaa-uni.edu.vn}
\affiliation{Faculty of Materials Science and Engineering, Phenikaa University, Hanoi 12116, Vietnam}
\affiliation{Phenikaa Institute for Advanced Study (PIAS), Phenikaa University, Hanoi 12116, Vietnam}

\maketitle
To validate the modified WHEP model, we apply Eq.(12) to representative ionic crystals, including LiF, FeO, CaO, CsI, KBr, and KCl. For simplicity, equation-of-state parameters are extracted from recent DAC experiments (see Table S1). Figure S1 shows a quantitative consistency between our one-phase approach and available static/dynamic measurements up to 100 GPa. Hence, Eq.(12) can serve as a useful tool to investigate high-pressure melting processes.

\begin{table}[htp]
\renewcommand{\thetable}{S\arabic{table}}
\begin{ruledtabular}
\begin{tabular}{cccccc}
\textrm{Material}&
\textrm{Phase}&
\textrm{$K_T(0)$ (GPa)}&
\textrm{$K_T^{'}(0)$}&
\textrm{$T_m(0)$ (K)}&
\textrm{Reference}\\
\colrule
LiF&B1&66.2 &4.6 &1118 &[S1, S2]\\
FeO&B1&149.4 &3.6&1650&[S3, S4]\\
CaO&B1&111 &4.2&3222&[S5, S6]\\
CsI&B2&11.9 &5.8 &899&[S7, S8]\\
KBr&B2&14.9 &5.8 &881\footnote{This value is determined by fitting Eq.(12) to the B1-B2-liquid triple point in Ref.[S10].}&[S9, S10]\\
KCl&B2&17.2 &5.9 &$921^{a}$&[S9, S10]\\
\end{tabular}
\caption{\label{tab:table1} Equation-of-state parameters and the initial melting temperature of LiF, FeO, CaO, CsI, KBr, and KCl.}
\end{ruledtabular}
\end{table}

\begin{figure}[htp]
\renewcommand{\thefigure}{S\arabic{figure}}
\includegraphics[width=14.1 cm]{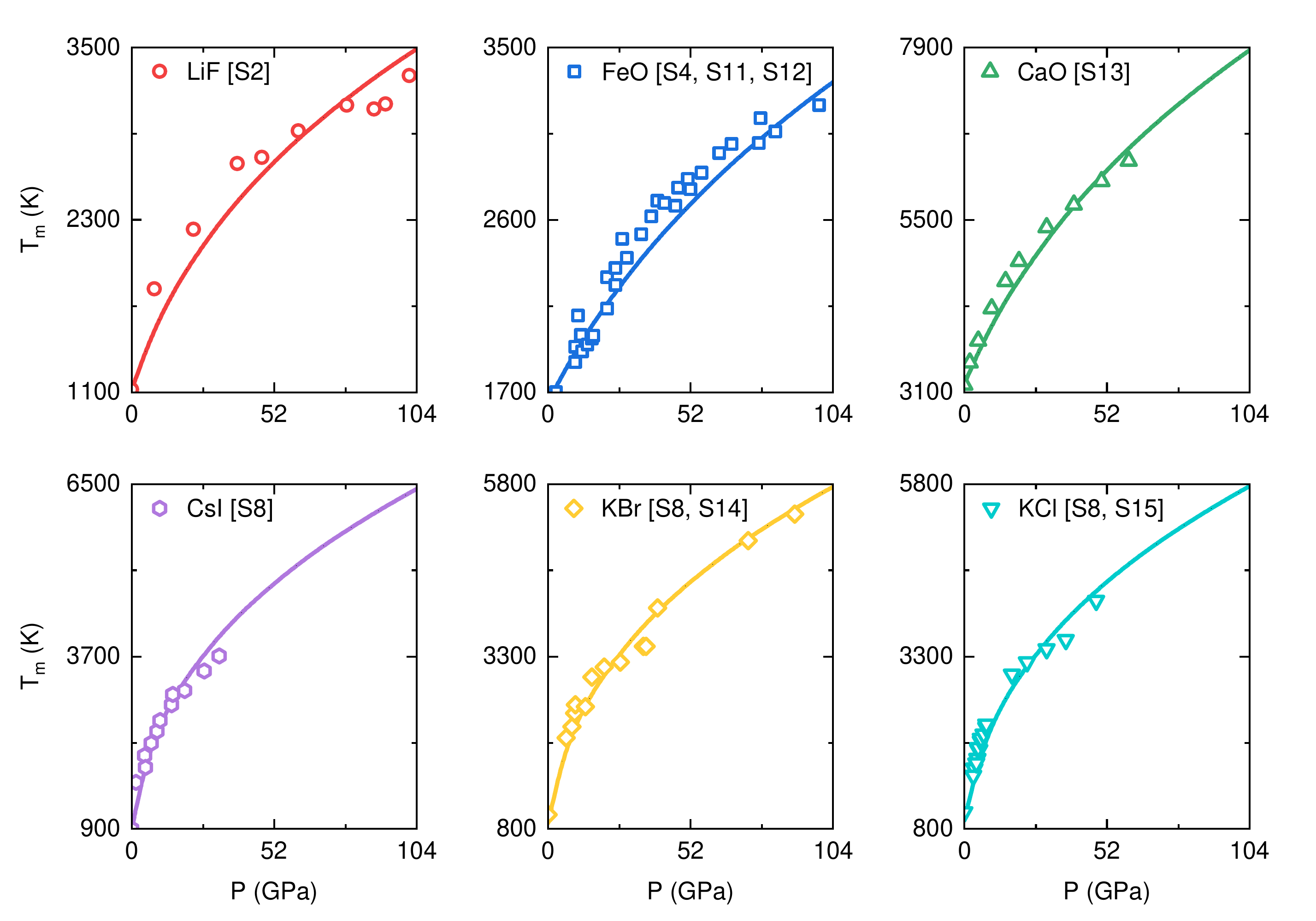}
\caption{(Color online) High-pressure melting profiles of typical ionic substances given by the modified WHEP theory (solid line) and previous measurements (open point). For CaO, data points denote simulation results due to the lack of experiments.}
\end{figure}

